\documentclass[letterpaper, 10 pt, conference]{ieeeconf}  % Comment this line out if you need a4paper

\IEEEoverridecommandlockouts                              % This command is only needed if 
\usepackage{color,xcolor}
\usepackage{graphicx}   
\usepackage{mathptmx} % assumes new font selection scheme installed
\usepackage{amsmath} % assumes amsmath package installed
\usepackage{amssymb}  % assumes amsmath package installed
\usepackage{tabularx}
\usepackage{hyperref}

\title{\LARGE \bf
Unified controller for take-off and landing for a fixed-wing aircraft
}

\author{Andres Montes de Oca and Gerardo Flores% <-this % stops a space
\thanks{*This work was supported in part by the FORDECYT-CONACYT under grant 292399.}% <-this % stops a space
\thanks{Andres Montes de Oca and Gerardo Flores are with the perception and Robotics Laboratory at Centro de Investigaciones en \'{O}ptica, Le\'{o}n, Guanajuato, Mexico, 37150. Email: {\tt\small andresmr@cio.mx, gflores@cio.mx}}%
}

\begin{document}

\maketitle
\thispagestyle{empty}
\pagestyle{empty}

%%%%%%%%%%%%%%%%%%%%%%%%%%%%%%%%%%%%%%%%%%%%%%%%%%%%%%%%%%%%%%%%%%%%%%%%%%%%%%%%
\begin{abstract}
Take-off and landing are the most important maneuvers for an aircraft's flight. Deployment for small fixed-wing aircraft is usually made by hand but when payload increases, take-off, and landing maneuvers are then performed on a runway making the procedures more complex. For that reason, we address the performance of the two maneuvers in order to develop a unique controller using the feedback control technique. We present the longitudinal aircraft dynamics to model the take-off and landing considering the rolling resistance forces during ground roll through a friction model. We also present the controller design for such a model. A stability proof is conducted to validate the stability of the system with the developed control law. Additionally, simulations are carried out to corroborate that the control law is effective applied to the dynamic model presented.
\end{abstract}

%%%%%%%%%%%%%%%%%%%%%%%%%%%%%%%%%%%%%%%%%%%%%%%%%%%%%%%%%%%%%%%%%%%%%%%%%%%%%%%%
\section{INTRODUCTION}
%small intro
Due to the exponential growth of the application of small drones, several control problems have emerged \cite{8453426}. In particular, the autonomy of take-off and landing for fixed-wing aircraft are not easy procedures to perform, mainly due to the fact that both tasks are integrated by several phases with important constrains to be considered. Since both procedures are important to successfully complete a mission, it is important to develop a robust controller to neglect ground effects and wind disturbances. For that, linear controllers have been developed, as well as nonlinear approaches such as fuzzy, sliding mode-based and adaptive controllers, among others \cite{7152420}, \cite{6564789}, \cite{6842272}. In this work we present a unified controller for take-off and landing maneuvers considering the rolling friction force caused by the contact with the runway.

%state of the art
Take-off for fixed-wing aircraft has not been investigated strongly unlike vertical take-off for such an aircraft as in \cite{7986093}, \cite{8821804}, \cite{8564504}, \cite{8821204}, \cite{CAKICI2016267}, \cite{LUGOCARDENAS2014713}, and \cite{8577478}. Within the existing study of take-off and landing for fixed-wing aircraft, few approaches have been proposed for the take-off problem as will be mentioned next. In \cite{7984131} and \cite{7984132}, real-time control is performed in the presence of windshears where a feedback strategy is used on the linearized dynamics of the aircraft model. In \cite{7838846}, it is designed a safe and robust automatic take-off maneuver. 
\begin{figure}[ht]
\begin{center}
\includegraphics[width=8.6cm]{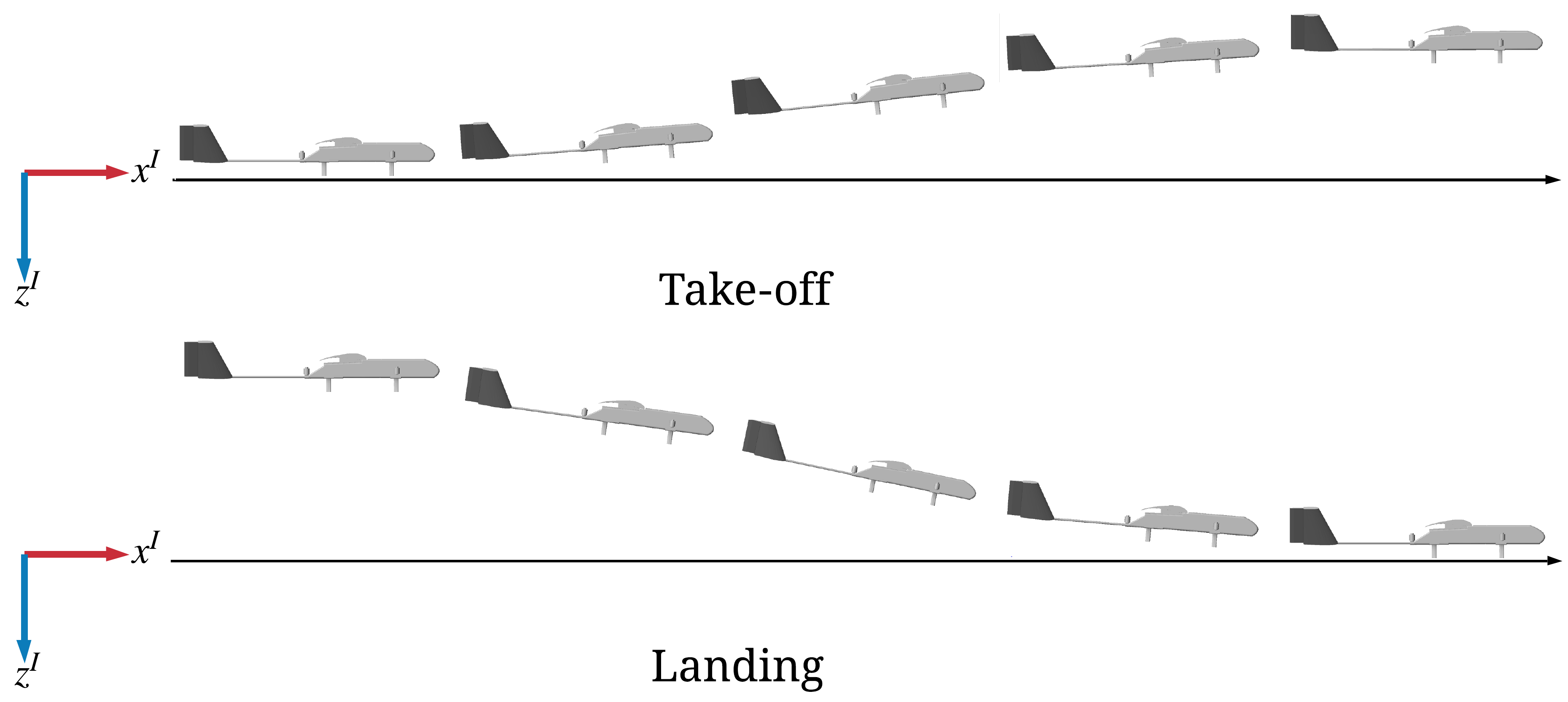}    % The printed column width is 8.4 cm.
\caption{3D simulation of take-off and landing maneuvers. The video showing the simulation results is available at \url{https://youtu.be/45nLCa8ALbE}.}
\label{fig:scheme}
\end{center}
\end{figure}
In \cite{7064151}, a dynamic model is developed involving the lateral friction generated by the wheels during take-off; a controller is also developed with a PID and fuzzy approach. On the other hand, landing has been addressed more as described next. In \cite{adaptive_landing}, it is developed a fully adaptive algorithm for autonomous landing. With the same approach, in \cite{autonomous_adaptive_landing}, an autonomous adaptive control system for the longitudinal model is developed using a neural network that provides the adaptive component of the control law. In \cite{longitudinal_landing}, a low-cost autonomous runway landing control system for fixed-wing is developed through a PID design method. Another technique is used in \cite{safe_landing}, where a Linear Quadratic Regulator (LQR) controller that accurately lands the vehicle on a runway is used to compute the safety region to determine safe landing. To minimize the control law switching, in \cite{robust_landing}, automatic landing is also addressed using classical loop-shaping and robust control techniques are used to design the individual control loops. Another techniques are used for autonomous landing performance as is the tracking model predictive static programming guidance and dynamic inversion for the outer and inner loop, respectively in \cite{autonomous_landing}. Adding difficult atmospheric conditions like crosswind and turbulence, autonomous landing is achieved through longitudinal and lateral controller designs as in \cite{autonomous_atmospheric_l}. In \cite{adaptive_flare}, it is proposed an adaptive landing scheme towards bounded variations in flight to velocity, altitude, and flight path angle. Following the control switching approach, in \cite{accurate_landing}, it is developed a control system architecture with strong disturbance rejection using a combination of controllers at different stages during the landing phases. Both maneuvers are controlled as in \cite{l1_landing} where a high performance $L_{1}$ adaptive control method is proposed, based on a combination of $L_{1}$ adaptive controller and a robust pole assignment controller. The latter is used to stabilize the inner close-loop system while the $L_{1}$ adaptive controller is used to reject disturbances and unknown ground effects. In \cite{take_off_landing_algorithms}, a take-off and landing algorithm has been developed and implemented on a low-cost flight control system using a PID approach. A comparison of flight controls for landing is made including conventional PID and neural net-based and fuzzy logic-based controls applied to the longitudinal model in \cite{comparison}.  

%contribution
Despite the similarities between take-off and landing, both tasks have been treated as different processes each controlled separately by different controllers at their different stages. We propose a unified controller for both take-off and landing maneuvers for a fixed-wing aircraft, reducing the number of applied controllers along with both procedures. We analyze the longitudinal aircraft dynamics for both maneuvers considering the rolling forces during the first phase of the take-off and also during the last phase of landing to develop a control law. A Lyapunov stability analysis is also presented showing local asymptotic stability as well as simulation results for both flight stages.

%organization
The paper is organized as follows. In Section \ref{sec:problem_statement} we present the description of the processes we want to control: take-off and landing maneuvers; we also introduce the problem statement. In Section \ref{sec:main_results}, the main results are presented as is the theorem related to the control law we developed and also the proof of the theorem. Section \ref{sec:simulation} shows simulations related to the control law applied to the system under take-off and landing schemes. Finally, in Section \ref{sec:conclusion}, a conclusion is made, summing up the conducted work and mentioning future work.

\section{System description and problem statement}\label{sec:problem_statement}

\subsection{Take-off and landing maneuvers' description}\label{subsec:description}
The take-off is the first stage of the flight which allows the aircraft to become airborne and start a flight trajectory. After the completion of the aircraft's mission, it should go back to the ground performing a smooth landing as the ending stage of the flight. The take-off and landing are very important for the aircraft operation because the majority of accidents occur during these two maneuvers, as it is explained in \cite{performance}. Both maneuvers are divided into several stages in which the aircraft switches between them according to its current velocity. It is important to mention that such a velocity of the aircraft is relative to the surrounding air and is known as the airspeed vector 
\begin{equation}
    \vec{V} = [u, w]^{T}
    \label{eq:V_a}
\end{equation}
with velocity components of the aircraft ($u$, $w$) relative to the inertial frame expressed in the body-fixed frame ($x^{I}$, $z^{I}$). The airspeed vector's magnitude is computed as follows \cite{small_unmanned}
\begin{equation}
    V = \sqrt{u^{2}+w^{2}},
    %\label{eq:airspeed}
\end{equation}
but since $w \ll u$, one can consider the velocity in the longitudinal plane to be \cite{u_velocity}
\begin{equation}
    V \approx u.
    \label{eq:airspeed}
\end{equation}
Such airspeed must also be equal to the stall velocity that is the minimum velocity required for the aircraft to maintain flight level with zero acceleration. The stall velocity can be computed with the following expression \cite{stall_speed}
\begin{equation}
V_{stall}=\sqrt{\frac{2 mg}{\rho C_{L_{max}}S}}
\label{eq:v_stall}
\end{equation}
where $\rho$ is the air density; $S$ is the wing surface area; $C_{L_{max}}$ is the maximum lift coefficient; $m$ is the aircraft mass; and $g$ is the gravity constant. It is important to note that the phases involved in the take-off and landing maneuvers highly depend on the stall velocity. This is due to the fact that maintaining this velocity helps the aircraft to stay airborne. 
\begin{figure}[ht]
\begin{center}
\includegraphics[width=8.6cm]{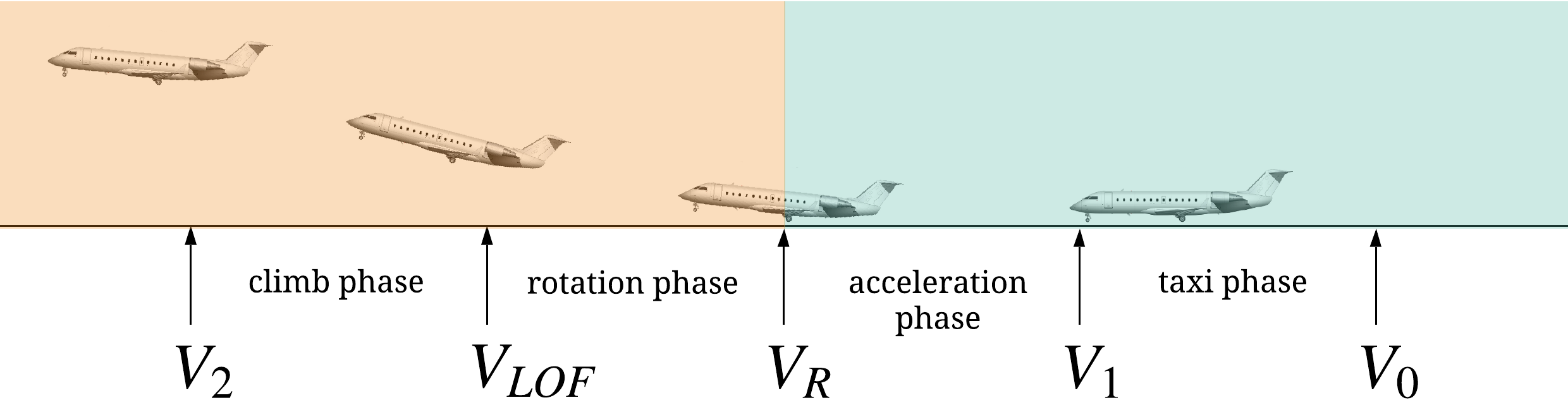}    % The printed column width is 8.4 cm.
\caption{Phases during take-off. Light red background indicates airborne phases and light blue background indicates ground phases. During ground phases, the friction model \eqref{eq:friction_model} is valid.}
\label{fig:takeoff_phases}
\end{center}
\end{figure}

The take-off task for a fixed-wing can be divided into the following main phases shown in Fig. \ref{fig:takeoff_phases} \cite{take_off_phases}:
\begin{enumerate}
    
    \item \textit{Taxi}. The aircraft starts accelerating from zero velocity $V_{0}$ to an arbitrary low ground velocity $V_{1}$. Thrust is used to ensure the aircraft to maintain low ground speed. Here, the pitch is maintained at zero.
    \item \textit{Acceleration}. The aircraft starts accelerating at full throttle until it reaches an appropriate take-off rotation velocity $V_{R}$ for the next phase. Such velocity is computed as follows
\begin{eqnarray}
    V_{R} = 1.1 (V_{stall})\label{eq:V_R}.
\end{eqnarray}
During this phase, the pitch angle is still maintained at zero.

\item \textit{Rotation}. In this phase, the aircraft's pitch angle should start increasing gradually up to a maximum pitch angle. As a consequence, the aircraft gains altitude passing the ground level. Here, the front portion of the aircraft is generally lifted first, and soon after the rear portion is also lifted to make the aircraft airborne \cite{performance}. During a small period, rolling resistance forces still affect the aircraft dynamics while part of it is in contact with the ground. At the end of this phase, the aircraft must reach the lift-off velocity that satisfies the following expression
\begin{equation}
    V_{LOF}=1.15(V_{stall}).
\end{equation}

\item \textit{Climb}. Once the aircraft is above the ground, it must keep the pitch increment so it reaches the maximum pitch angle. After that, the aircraft should decrease the pitch to a desirable value that allows it to maintain a constant altitude. At this point, rolling forces are no longer affecting the dynamics of the aircraft. Once climb velocity is reached, constant altitude must be maintained through an altitude control technique. The velocity of the aircraft during this phase is called climb velocity $V_{2}$ and must satisfy the following relation
\begin{equation}
    V_{2}=1.2(V_{stall}).\label{eq:V2}
\end{equation}
\end{enumerate}

On the other hand, landing is divided into the following phases as is shown in Fig. \ref{fig:landing_phases} \cite{phases}:
\begin{figure}[ht]
\begin{center}
\includegraphics[width=8.6cm]{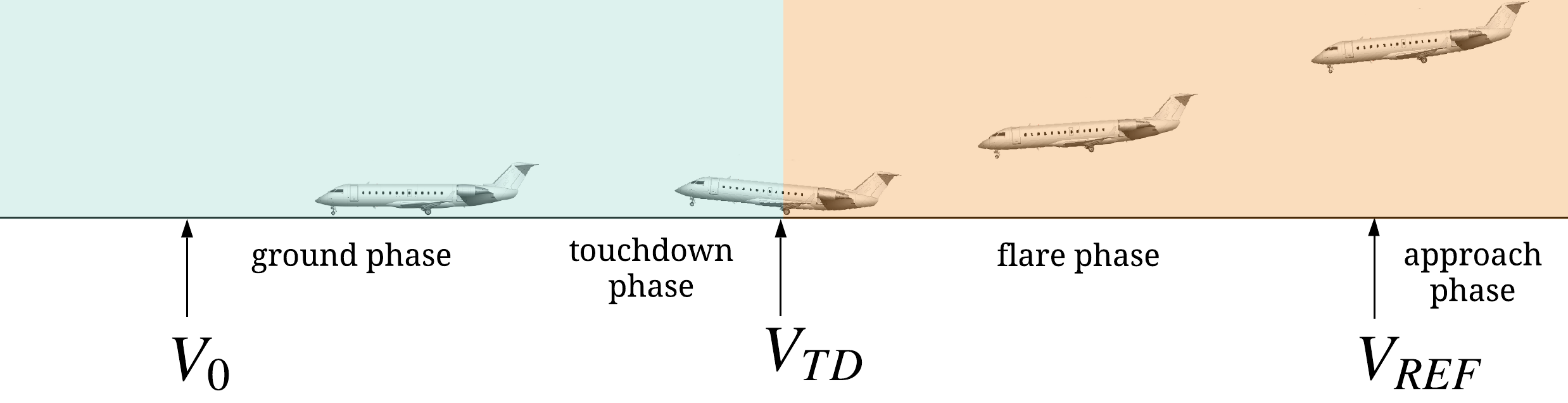} 
\caption{Phases of landing. Light red background indicates the airborne phases and light blue background indicates the ground phases. During ground phases, model \eqref{eq:friction_model} is valid.}
\label{fig:landing_phases}
\end{center}
\end{figure}
\begin{enumerate}
    \item \textit{Approach}. In this phase, the aircraft must decrease its pitch angle up to a minimum angle. The aircraft also starts decreasing its velocity, reaching the approach airspeed $V_{REF}$ that must satisfy the following relation
    \begin{equation}
        V_{REF}= 1.3(V_{stall})\label{eq:V_D}.
    \end{equation}
    
    \item \textit{Flare}. Once the aircraft reaches the approach velocity $V_{REF}$, it should continue decreasing its velocity to the touchdown velocity $V_{TD}$. Such a velocity must satisfy the following 
    \begin{equation}
        V_{TD}=1.1(V_{stall}).\label{eq:V_TD}
    \end{equation}
    
    \item \textit {Touchdown}. At this point, the aircraft makes contact with the ground and rolling forces start affecting the aircraft dynamics. Here, the pitch angle is zero.
    
    \item \textit{Ground}. The remaining acceleration allows the vehicle to move along the runway until it reaches zero velocity $V_{0}$ or brakes are activated.
\end{enumerate}

As seen, some of the phases of the take-off and landing involve contact with the runway, so friction force is inherent to both processes. Such force acts on the aircraft dynamics due to the contact of the aircraft's wheels with the runway's surface. This effect is known as rolling resistance \cite{rolling_def}.

\subsection{Modeling}\label{subsec:modeling}
\begin{figure}[ht]
\begin{center}
\includegraphics[width=8.6cm]{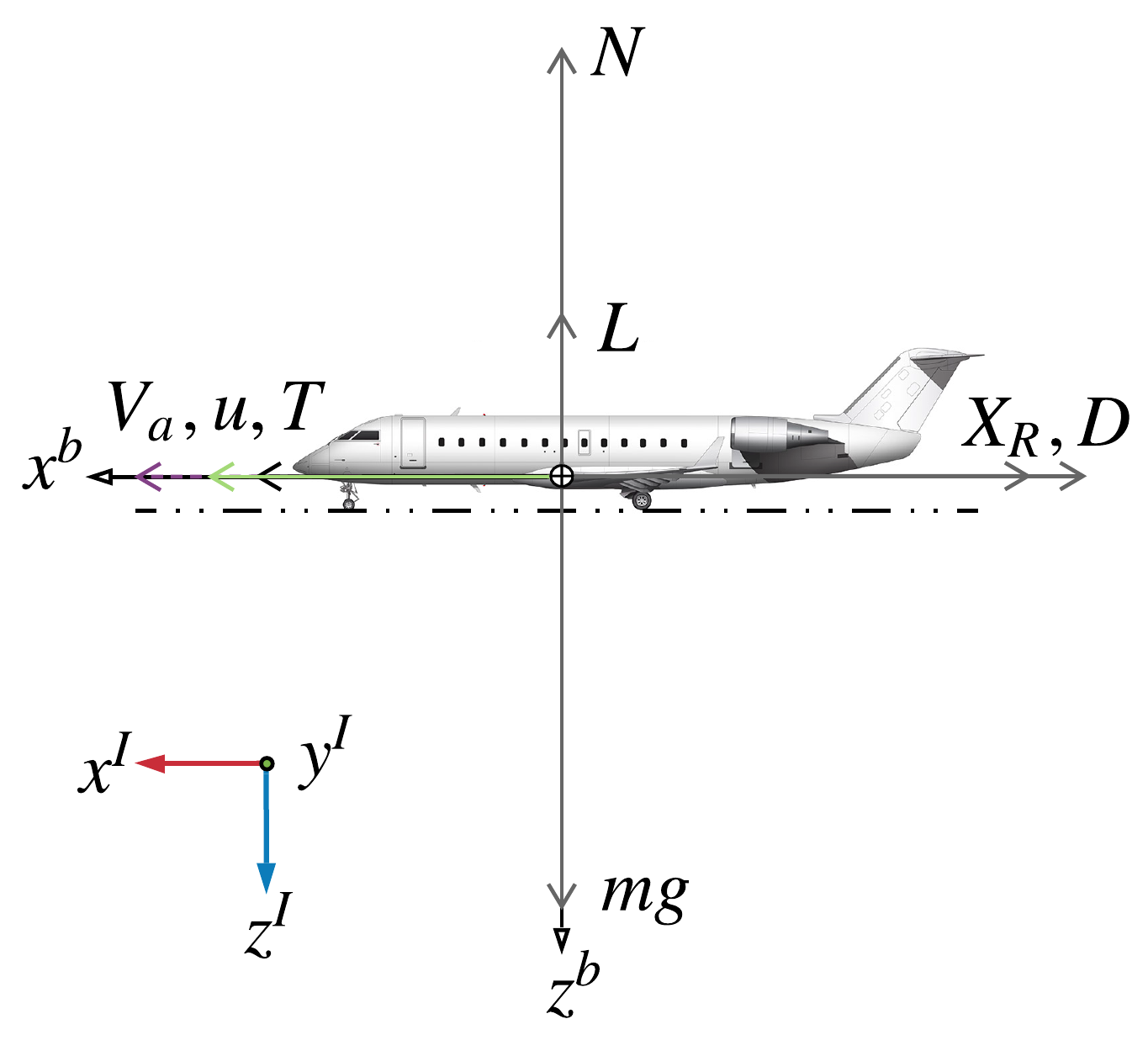}
\caption{Longitudinal model considering the rolling resistance forces during ground phases. The dotted line indicates ground level.}
\label{fig:ground_effect}
\end{center}
\end{figure}
\begin{figure}[ht]
\begin{center}
\includegraphics[width=8.4cm]{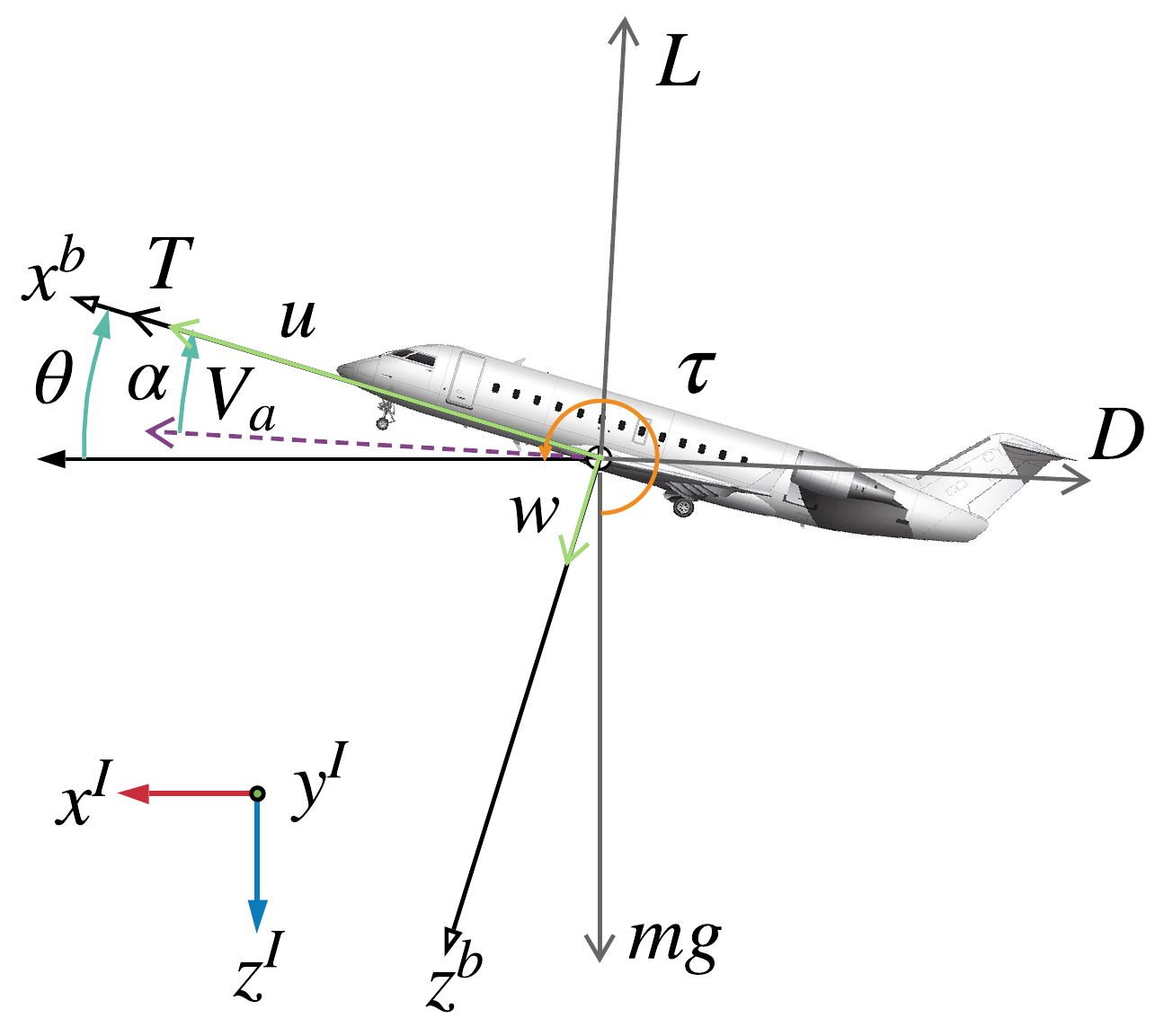}    % The printed column width is 8.4 cm.
\caption{Longitudinal model of the fixed-wing aircraft.}
\label{fig:model}
\end{center}
\end{figure}

 
In this work we take the longitudinal aircraft dynamics \cite{model} adding rolling resistance forces. Let consider Figs. \ref{fig:ground_effect} and \ref{fig:model} as reference for the following analysis. We take position and attitude dynamics from \eqref{eq:dynamics} and introduce the rolling resistance force generated due to the ground roll, as in \cite{canard}. The rolling resistance force component $X_{R}$ is expressed in the body frame relative to the inertial frame along the $x^{I}$ axis as can be seen in Fig. \ref{fig:ground_effect} and is equivalent to the following expression \cite{friction_force}
\begin{equation}
    \mu N = \mu ( L \cos(\theta - \alpha) - D \sin(\theta - \alpha) +T \sin(\theta)- mg)
\end{equation}
where $\mu$ is the coefficient of friction of the ground; $N$ is the normal force expressed in the $z^{I}$ axis; $L$ is the lift force; $D$ is the drag force; $\theta$ is the pitch angle; and $\alpha$ is the angle of attack.
So, let define the following system
\begin{equation}
\begin{aligned}
\label{eq:dynamics}
%\label{eq:dyn_position}
        \dot{u} =& - qw - g\sin \theta + a[L \sin \alpha -D \cos \alpha + T \\
        &-\mu ( L \cos(\theta - \alpha) - D \sin(\theta - \alpha) +T \sin(\theta) \\
        &- mg)]\\
        \dot{w} =& qu + g\cos \theta + a(-D \sin \alpha - L \cos \alpha)\\
        \dot{\theta} =& q\\
        \dot{q} =& \tau
\end{aligned}
\end{equation}
where ($u$, $w$) are the velocity components of the aircraft relative to the inertial frame expressed in the body-fixed frame ($x^{I}$, $z^{I}$); $q$ is the pitch rate; $T$ is the thrust force generated by the aircraft's engine; $\tau = \frac{M_{m}}{J_{yy}}$ is the torque that encloses the aerodynamic moment $M_{m}$ and the inertia matrix of the aircraft $J_{yy}$, around the inertial $y^{I}$ axis. ($T, \tau$) will be later treated as control inputs. The lift and drag forces can be computed with the following expressions \cite{small_unmanned}
\begin{equation}
\begin{aligned}
L = K V^{2} C_{L}
\\
D = K V^{2} C_{D} 
\end{aligned}\label{eq:D_L}
\end{equation}
where $K= \frac{\rho S}{2}$; $C_{L}$ and $C_{D}$ are the lift and drag coefficients at the desired angle of attack $\alpha$.

To model the presence of the friction coefficient we use the following function
\begin{eqnarray}
\label{eq:friction_model}
F_{\mu}(t)=&\begin{cases}
        	0 \hspace{13pt} $if$ \hspace{10pt} N \geq 0\\
          \mu  \hspace{12pt} $if$ \hspace{10pt} N < 0.
	\end{cases}
\end{eqnarray}
Note that when the aircraft becomes airborne, the sum of the force components of the aircraft along the $z^{I}$ axis is equal or greater than the weight of the aircraft. Under this condition, the friction should not exist.

\subsection{Problem statement}\label{subsec:PE}
The rolling resistance forces due to friction generated by the wheels during taxi and touchdown phases are not always considered for the design of the controllers. The aim of this work is to provide a unique simple controller with control inputs ($T$, $\tau$) for both take-off and landing maneuvers for a fixed-wing aircraft. It is also important to develop a model for the resistance coefficient ($\mu$) during rolling. In this work, we address the problem with the longitudinal aircraft dynamics, developing a control law for such a system to achieve a proper take-off and landing. 

\section{Main results}\label{sec:main_results}

\subsection{Design of the desired trajectories}\label{subsec:trajectories}
We now explain the computation of the desired trajectories for $(u_{d}, \theta_{d}, q_{d})$ to ensure the aircraft to perform an adequate take-off and landing. 

\subsubsection{Take-off}\label{sssec:takeoff}
As depicted in Fig. \ref{fig:takeoff_phases}, the required airspeed for take-off should behave in an increasing way through the following set
\begin{equation}
V_{d}^{T} = 
\begin{Bmatrix}
 V_{0} &=& 0\\ 
 V_{1} &=& 0.5 (V_{stall})\\ 
 V_{R} &=& 1.1 (V_{stall})\\
 V_{LOF} &=& 1.15 (V_{stall})\\
 V_{2} &=& 1.2 (V_{stall})
\end{Bmatrix}.
\label{eq:takeoff_velocities}
\end{equation}
Note that for $V_{1}$ which must be an arbitrary value between zero velocity and $V_{R}$ we choose half the stall velocity. Then, we compute $u_{d}$ from \eqref{eq:airspeed}, so we can state that the desired value for $u_{d} \approx V_{d}$ where $v_{d}$ takes the previous set \eqref{eq:takeoff_velocities}.

The desired trajectory that must be followed by the pitch angle should be an increasing peak from zero to a maximum value to allow the rotation of the aircraft. Such a value is usually up to $15^{\circ}$ \cite{take_off_phases}. After reaching the maximum pitch angle, it should decreases to a defined value in order to keep climbing or maintain flight level. We propose the use of the following Gaussian function to obtain the trajectory for the desired pitch angle
\begin{equation}
    \theta_{d}= \theta_{lim} e^{\big(\frac{-0.5(V_{d}-c)^2}{d^2}\big)}\label{eq:theta_function}
\end{equation}
that is dependant on the airspeed; ($c$, $d$) are positive real numbers that can be adjusted to achieve the pitch increment according to the rotation velocity; $\theta_{lim}$ is the maximum value for the pitch angle. The pitch rate trajectory is obtained from the time derivative of $\theta_{d}$ as follows 
\begin{equation*}
    q_{d}  = \frac{d}{dt}\theta_{d}.
\end{equation*}

\subsubsection{Landing}\label{sssec:landing}
For landing, Fig. \ref{fig:landing_phases} provides the decreasing airspeed as is expressed in the following set
\begin{equation}
V_{d}^{T} = 
\begin{Bmatrix}
 V_{REF} &=& 1.3(V_{stall})\\ 
 V_{TD} &=& 1.1(V_{stall})\\ 
 V_{0} &=& 0
\end{Bmatrix}.
\label{eq:landing_velocities}
\end{equation}
As $u_{d} \approx V_{d} $, the desired trajectory for $u_{d}$ will be then composed by the set \eqref{eq:landing_velocities}.

The trajectory for the pitch angle should consider a decreasing peak with a minimum pitch angle that allows the aircraft to lose altitude enough to stay close to the ground. Then, the pitch angle should increase from that minimum value to zero in order to make the aircraft land and follow the runway. Such a minimum value is not specified as this may vary according to the aircraft and landing strategy. For that, we can use \eqref{eq:theta_function} to obtain the necessary trajectory. The pitch rate can be obtained as for the take-off, through the time derivative of the desired pitch angle.

\subsection{Control algorithm}
Within the proposed control law we use a saturated function specifically in the control input $T$, so it is important to introduce it as well as its properties. Consider the following Definition.
\newline
\textit{Definition 1.}
Given two positive constants $L$, $M$ with $L \leq M$, a function $\sigma: \mathbb{R} \rightarrow \mathbb{R}$ is said to be a linear saturation for ($L$, $M$) if it is a continuous and non-decreasing function satisfying the following conditions.
\begin{enumerate}
    \item $s\sigma(s)>0 \hspace{5pt} \forall \hspace{5pt} s \neq 0$
    \item $\sigma(s)=s$ when $|s|\leq L$
    \item $|s\sigma(s)|\geq M  \hspace{5pt} \forall \hspace{5pt} s\in \mathbb{R}$\\
\end{enumerate}
%\end{definition}
Now let consider the saturation function ($\sigma$) \cite{8619303}
\begin{equation}
    \sigma(s) =\begin{cases}
        \frac{\tan^{-1}(n(s-L))}{n}+L \hspace{15pt}$if$\hspace{15pt} s>L\\
        \frac{\tan^{-1}(n(s+L))}{n}+L \hspace{15pt}$if$\hspace{15pt} s<-L\\
        s \hspace{75pt}$if$\hspace{12pt} |s|\leq L
        \end{cases}\label{eq:saturation}
\end{equation}
where 
\begin{equation*}
    n = \frac{\pi}{2(M-L)}.
\end{equation*}
Now that the saturated function is defined, we describe the control law as follows. 

\textit{Theorem 1.}
Consider the system \eqref{eq:dynamics} with the control inputs ($T$, $\tau$) defined by
\begin{equation}
\label{eq:control_functions}
\begin{aligned}
         T=& -k_{T}\sigma(u-u_{d})+\frac{1}{a}\Big[ \sin \theta+KV^{2}(C_{D}\cos \alpha -C_{L}\sin \alpha) \\
         &+\mu N+qw+\dot{u}_{d}\\
         \tau=& -k_{\theta} (\theta-\theta_{d})-k_{q}(q-q_{d}) + \dot{q}_{d}\Big]
         \end{aligned}
\end{equation}
where ($k_{T}$, $k_{\theta}$, $k_{q}$) are positive real numbers; ($u_{d}$, $\theta_{d}$, $q_{d}$) are the desired trajectories for the $u$ velocity component in the body-fixed frame, and also for pitch, and pitch rate, respectively. Then, the system with the above control law is locally asymptotically stable, making the aircraft to perform take-off and landing by converging ($u$, $\theta$, $q$) to desired trajectories ($u_{d}$, $\theta_{d}$, $q_{d}$) when $t \rightarrow \infty$.
%\end{thm}

\begin{proof}    % and the pf environment for proofs
Let define the variable change 
\begin{equation}
\begin{aligned}
\label{eq:error_dynamics}
e_{1} & = u- u_{d}\\ 
%e_{2} & = w- w_{d}\\ 
%e_{2} & = \dot{w} - \dot{w}_{d}\\
e_{2} & = \theta- \theta_{d}\\ 
e_{3} & =  q - q_{d}.\\
\end{aligned}
\end{equation}
Then to prove that the errors ($e_{1}$, $e_{2}$, $e_{3}$) converge to zero, we propose the positive definite Lyapunov Candidate Function
\begin{equation*}
    V(e_{1}, e_{2}, e_{3}) = \frac{1}{2} e_{1}^{2} + \frac{1}{2} e_{2}^{2} + \frac{1}{2} e_{3}^{2}
\end{equation*}
whose time derivative is 
\begin{equation}
        \dot{V}(e_{1}, e_{2}, e_{3})=e_{1} \dot{e}_{1} + e_{2} \dot{e}_{2} + e_{3} \dot{e}_{3}\label{eq:v_dot}
\end{equation}
and for which we need to prove that:
\begin{enumerate}
    \item $\dot{V}(e_{1}, e_{2}, e_{3}) < 0 \hspace{5pt} \forall \hspace{5pt} e_{1},e_{2},e_{3} \neq 0$ 
    \item $\dot{V}(0) = 0$. \label{eq:condition}
\end{enumerate}
Substituting \eqref{eq:error_dynamics} and \eqref{eq:dynamics} in \eqref{eq:v_dot}, we obtain
\begin{equation*}
\begin{aligned}
    \dot{V}(e_{1}, e_{2}, e_{3}) =& e_{1} [- qw -g \sin(\theta) + a(L\sin(\alpha) - D\cos(\alpha) \\
    &+ T - \mu N - \dot{u}_{d})] + e_{2} e_{3} + e_{3} (\tau- \dot{q}_{d}) 
\end{aligned}
\end{equation*}
Then, substituting the control inputs ($T$, $\tau$) from \eqref{eq:control_functions} we have the simplified function
\begin{equation*}
\begin{aligned}
    \dot{V}(e_{1}, e_{2}, e{3}) =& -k_{T}e_{1}\sigma(e_{1}) + e_{2} e_{3} + e_{3} (-k_{\theta}e_{2}-k_{q}e_{3}) 
\end{aligned}
\end{equation*}
\begin{equation}
\begin{aligned}\label{eq:lyapunov_separated}
        \dot{V}(e_{1}, e_{2}, e_{3})= \underbrace{-k_{T}e_{1}\sigma(e_{1})}_{\dot{V}_1} \underbrace{- k_{q}e_{3}^{2}}_{\dot{V}_{2}}\underbrace{+e_{2}e_{3}(1- k_{\theta})}_{\dot{V}_{3}}
\end{aligned}
\end{equation}
in which it is clear that $\dot{V}(0) = 0$ holds. Now to prove that $\dot{V}(e_{1}, e_{2}, e_{3}) < 0 \hspace{5pt} \forall \hspace{5pt} e_{1},e_{2},e_{3} \neq 0$, we divide \eqref{eq:lyapunov_separated} into three terms ($\dot{V}_{1}$, $\dot{V}_{2}$, $\dot{V}_{3}$). If we consider the condition (1) from Definition 1, is clear that $\dot{V}_{1}$ is always negative. The second term $\dot{V}_{2}$ that contains a quadratic term will always be negative either $e_{3}$ is negative of positive. Finally, the term $\dot{V}_{3}$ will only be positive when $e_{2}e_{3} < 0$. In this specific case, only when $e_{2}< e_{3}$, $\dot{V}_{3} < -\dot{V}_{2}$ making $\dot{V}$ negative. With this, it is demonstrated local asymptotically stability.
\end{proof}

\section{Simulations}\label{sec:simulation}
Simulations were carried out to verify the effectiveness of the designed control law. We present simulations for take-off and landing by separate.
The parameters related to the aircraft that were used in the simulation, as well as the stall velocity from \eqref{eq:v_stall}, are shown in Table \ref{tab:Parameters_aircraft} .In Table \ref{tab:Parameters_controllers} are shown the parameters and gains for the control inputs ($T$, $\tau$). As a reference, altitude is computed to observe its behavior during take-off and landing. That is done integrating the following expression
\begin{equation*}
    \dot{h}=-u\sin(\theta)+w\cos(\theta)
\end{equation*}
that corresponds to the velocity of the aircraft relative to the inertial frame along the $z^{I}$ axis.
\begin{table}[ht]
	%\centering
	\caption{Parameters of the aircraft.}
	\label{tab:Parameters_aircraft}
    \begin{tabularx}{\columnwidth}{p{0.5\columnwidth} p{0.5\columnwidth}}
			\hline\hline\\[-3mm]
			\multicolumn{1}{c}{\textbf{Parameter}} &\multicolumn{1}{c}{{\textbf{Value}}}  \\[1.6ex] \hline
			\multicolumn{1}{c}{$S$}			& \multicolumn{1}{c}{$2$ $[m^{2}]$}\\
			\multicolumn{1}{c}{$m$}	  		& \multicolumn{1}{c}{$3$ $[kg]$}\\
            \multicolumn{1}{c}{$C_{l_{max}}$}  	& \multicolumn{1}{c}{$1.25$}\\
            \multicolumn{1}{c}{$\rho$} & \multicolumn{1}{c}{$1.22$ $[kg/m^{3}]$}\\
            \multicolumn{1}{c}{$V_{stall}$} & \multicolumn{1}{c}{$4.39$ $[m/s]$}\\
            \multicolumn{1}{c}{$\mu$} & \multicolumn{1}{c}{$0.02$}\\
            %\multicolumn{1}{c}{$b$} & \multicolumn{1}{c}{$1$ $[kgm^{2}]$}\\
			\hline\hline
		\end{tabularx}
\end{table}
\begin{table}[ht]
	\renewcommand{\arraystretch}{1.3}
	\caption{Gains and parameter of the controllers.}
	\centering
	\label{tab:Parameters_controllers}
    \begin{tabularx}{\columnwidth}{p{0.5\columnwidth} p{0.5\columnwidth}}
			\hline\hline \\[-3mm]
			\multicolumn{1}{c}{\textbf{Parameter}} &\multicolumn{1}{c}{{\textbf{Value}}}  \\[1.6ex] \hline
            \multicolumn{1}{c}{$L$} & \multicolumn{1}{c}{$0.9$}\\
            \multicolumn{1}{c}{$M$} & \multicolumn{1}{c}{$1$}\\
            \multicolumn{1}{c}{$k_{T}$} & \multicolumn{1}{c}{$10$}\\
            \multicolumn{1}{c}{$k_{\theta}$} & \multicolumn{1}{c}{$3.3$}\\
            \multicolumn{1}{c}{$k_{q}$} & \multicolumn{1}{c}{$2$}\\
			\hline\hline
		\end{tabularx}
\end{table}

\subsection{Take-off}
\begin{figure}[ht]
\begin{center}
\includegraphics[width=8.4cm]{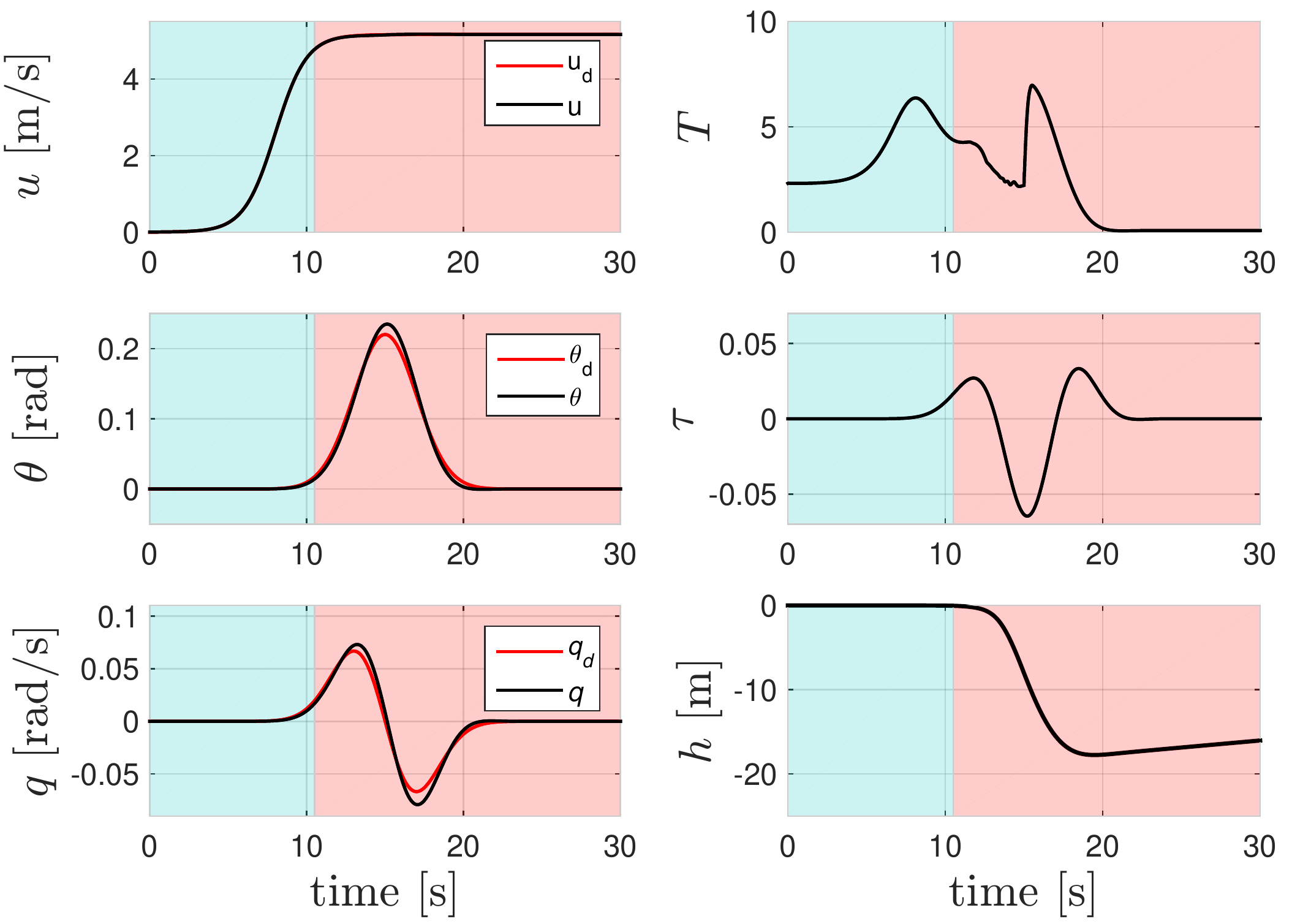} 
\caption{States and control inputs for take-off.} 
\label{fig:takeoff_plots}
\end{center}
\end{figure}

\begin{table}[ht]
	\renewcommand{\arraystretch}{1.3}
	\caption{Parameters and initial conditions for take-off.}
	\centering
	\label{tab:takeoff_values}
    \begin{tabularx}{\columnwidth}{p{0.5\columnwidth} p{0.5\columnwidth}}
			\hline\hline \\[-3mm]
			\multicolumn{1}{c}{\textbf{Parameter}} &\multicolumn{1}{c}{{\textbf{Value}}}  \\[1.6ex] \hline
            \multicolumn{1}{c}{$\theta_{lim}$} & \multicolumn{1}{c}{$0.22$ $[rad]$}\\
            \multicolumn{1}{c}{$c$} & \multicolumn{1}{c}{$2$}\\
            \multicolumn{1}{c}{$d$} & \multicolumn{1}{c}{$15$}\\
            \multicolumn{1}{c}{$u_{0}$} & \multicolumn{1}{c}{$0$ $[m/s]$}\\
            \multicolumn{1}{c}{$w_{0}$} & \multicolumn{1}{c}{$0$ $[m/s]$}\\
            \multicolumn{1}{c}{$\theta_{0}$} & \multicolumn{1}{c}{$0$ $[rad]$}\\
            \multicolumn{1}{c}{$q_{0}$} & \multicolumn{1}{c}{$0$ $[rad/s]$}\\
            \multicolumn{1}{c}{$h_{0}$} & \multicolumn{1}{c}{$0$ $[m]$}\\
			\hline\hline
	\end{tabularx}
\end{table}
Initial conditions and parameters used for the take-off simulation are shown in Table \ref{tab:takeoff_values}.
($u, \theta, q$) states and their desired trajectories as well as the control inputs ($T, \tau$), and altitude are shown in Fig. \ref{fig:takeoff_plots}. It can be seen that the states converge to the desired trajectories. Also, note that the altitude shown behaves as expected for the take-off. The negative altitude is due to the $z^{I}$ axis points downwards as can be seen in the diagram of Fig. \ref{fig:model}.
\subsection{Landing}
\begin{figure}
\begin{center}
\includegraphics[width=8.4cm]{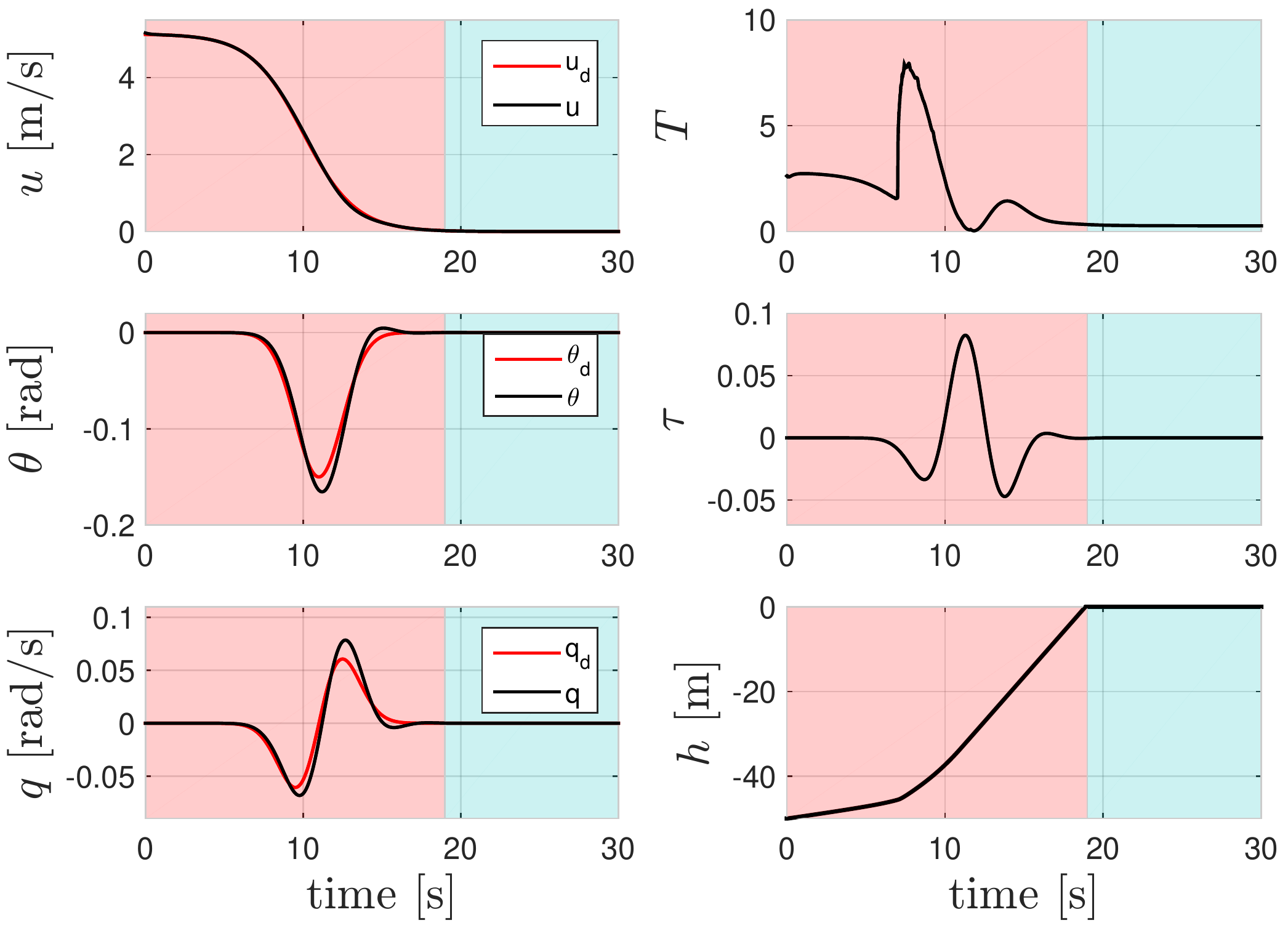}
\caption{States and control inputs for landing.} 
\label{fig:landing_plots}
\end{center}
\end{figure}

Parameters and initial conditions for landing can be seen in Table \ref{tab:landing_values}.
\begin{table}[ht]
	\renewcommand{\arraystretch}{1.3}
	\caption{Parameters and initial conditions for landing.}
	\centering
	\label{tab:landing_values}
    \begin{tabularx}{\columnwidth}{p{0.5\columnwidth} p{0.5\columnwidth}}
			\hline\hline \\[-3mm]
			\multicolumn{1}{c}{\textbf{Parameter}} &\multicolumn{1}{c}{{\textbf{Value}}}  \\[1.6ex] \hline
            \multicolumn{1}{c}{$\theta_{lim}$} & \multicolumn{1}{c}{$-0.15$ $[rad]$}\\
            \multicolumn{1}{c}{$c$} & \multicolumn{1}{c}{$1.5$}\\
            \multicolumn{1}{c}{$d$} & \multicolumn{1}{c}{$11$}\\
            \multicolumn{1}{c}{$u_{0}$} & \multicolumn{1}{c}{$5.16$ $[m/s]$}\\
            \multicolumn{1}{c}{$w_{0}$} & \multicolumn{1}{c}{$0$ $[m/s]$}\\
            \multicolumn{1}{c}{$\theta_{0}$} & \multicolumn{1}{c}{$0$ $[rad]$}\\
            \multicolumn{1}{c}{$q_{0}$} & \multicolumn{1}{c}{$0$ $[rad/s]$}\\
            \multicolumn{1}{c}{$h_{0}$} & \multicolumn{1}{c}{$-50$ $[m]$}\\
			\hline\hline
		\end{tabularx}
\end{table}
($u, \theta, q$) states and their desired trajectories, as well as control inputs ($T, \tau$), and altitude can be seen in Fig. \ref{fig:landing_plots}. It can be seen that the states converge to the desired trajectories. Also, note that altitude reaches ground level. The plot's background has been divided into two color sections to identify when the aircraft is in the air (light red) and when it has reached the ground (light blue). A video showing the 3D simulation of take-off and landing maneuvers can be found at the following link \url{https://youtu.be/45nLCa8ALbE}.
\section{Conclusion}\label{sec:conclusion}
Through the analysis of the rolling forces, that are involved from taxi to rotation phase during take-off, and from touchdown to ground phase, we aimed the integration of such forces into the longitudinal dynamics of the aircraft model. We also established a friction model that is added to the aircraft model, for which we developed a unique control law for both maneuvers in order to achieve a proper take-off and landing.  A stability proof for the control law is conducted and presented in this paper. To prove the effectiveness of the control law we perform simulations for the take-off and landing showing the convergence of the variable states to desired trajectories. For future work, we are encouraged to introduce moment terms involved due to the rolling forces, improve the desired trajectories, and expand this work for a 6 DOF dynamics model. We will also perform the implementation of our algorithms through the Hardware in the Loop technique with PX4 autopilot.


\bibliographystyle{IEEEtran}
\bibliography{g, IEEE_biblio}
\end{document}